\documentclass[prl,showpacs,twocolumn]{revtex4}
\usepackage{tabularx}
\usepackage{amsmath}
\usepackage{amssymb}
\usepackage{graphicx}
\usepackage{dcolumn}
\usepackage{bm}
\usepackage{ulem}
\usepackage{color}

\begin{document}

\title{Statistical Properties of Exciton Fine Structure Splittings and Polarization Angles in Quantum Dot Ensembles}
\author{Ming Gong$^{1}$}
\thanks{Present address: Department of Physics, The Chinese University of Hong Kong, Shatin, New Territories, Hong Kong, China} 
\thanks{Email: skylark.gong@gmail.com}
\author{B. Hofer$^2$}
\author{E. Zallo$^2$}
\author{R. Trotta$^{2,3}$}
\thanks{Email: rinaldo.trotta@jku.at}
\author{Junwei Luo$^{4}$}
\thanks{Email: jwluo.cn@gmail.com}
\author{Alex Zunger$^{5}$}
\author{O. G. Schmidt$^{2}$}
\author{Chuanwei Zhang$^{1}$}
\date{\today}

\begin{abstract}
We propose an effective model to describe the statistical properties of
exciton fine structure splitting (FSS) and polarization angle of quantum dot
ensembles (QDEs). We derive the distributions of FSS and polarization angle
for QDEs and show that their statistical features can be fully
characterized using at most three independent measurable parameters. The
effective model is confirmed using atomistic pseudopotential calculations as
well as experimental measurements for several rather different
QDEs. The model naturally addresses three fundamental questions that are
frequently encountered in theories and experiments: (I) Why the probability
of finding QDs with vanishing FSS is generally very small? (II) Why FSS and
polarization angle differ dramatically from QD to QD? and (III) Is there any
direct connection between FSS, optical polarization and the morphology of
QDs? The answers to these fundamental questions yield a completely new physical
picture for understanding optical properties of QDEs.
\end{abstract}

\affiliation{$^{1}$Department of Physics, The University of Texas at Dallas, Richardson,
TX, 75080 USA \\
$^{2}$Institute for Integrative Nanosciences, IFW Dresden, Helmholtzstr, 20,
D-01069 Dresden, Germany \\
$^{3}$Institute of Semiconductor and Solid State Physics, Johannes Kepler
University Linz, Altenbergerstr. 69 A-4040 Linz, Austria \\
$^{4}$National Renewable Energy Laboratory, Golden, Colorado 80401, USA \\
$^{5}$University of Colorado, Boulder, Colorado 80401, USA}

% 78.67.Hc  quantum dots
% 42.50.-p  Quantum optics
% 73.21.La  Quantum dots electron states and collective excitations in, 73.21.La
% 81.07.Ta  quantum dot fabrication
% The same pacs number is used in my previous prl paper  78.67.Hc, 42.50.-p, 73.21.La 
% G. Bester use 73.21.Hb, 42.50.-p, 73.21.La, 78.67.Hc 
% OKOKOKOK

\maketitle

%\thanks{Email: chuanwei.zhang@utdallas.edu}

\textit{Introduction.-} The fine structure splitting (FSS) of excitons in
self-assembled quantum dots (QDs) poses the major obstacle to the
realization of entangled photon pairs from biexciton cascade process, thus
has been a subject of extensive investigation in the past decade \cite%
{Akopian,Stevenson,Hofenbrak,Hudson,Young07,Young09,Gong08, WJP,Trotta}. 
Now it is quite clear that the FSS arises from the intrinsic
nonequivalence along [110] and [1$\bar{1}$0] directions in zinc-blende
crystals, which reduce the symmetry of the underlying lattice from $T_{d}$
to $C_{2v}$ for pure circular lens-shaped QDs, and the other nonuniform effects 
such as local strain, shape irregularities, alloys and interface effects \cite%
{Bester03, Bester05}, which further reduce the symmetry to $C_{1}$ for alloyed QDs
\cite{Gong11}. A single external field, such as electric field 
\cite{Kowalik,Gerardot,Vogel,Marcet,Luo12,Bennett}, magnetic field \cite{Stevenson, 
Stevenson06}, or anisotropic stress \cite{Plumhof,Singh10, Seidl, Chris, Luca,
Gong11}, is insufficient to eliminate the FSS because the lower bound of FSS
is generally much larger than the homogeneous broadening of the emission
line ($\sim $ 1 $\mu $eV). To eliminate the FSS, two non-equivalent
fields have to be combined \cite{WJP, Trotta}. Generally, the FSSs depend
strongly on the local details of QDs, thus is hard to be predicted in theories.
For instance for two QDs with the same morphology but tiny
difference in alloy atomistic arrangement, their FSS and polarization angle
may be fairly different. Therefore we confront three fundamental issues,
which are generally very challenging for understanding: (I) Why the
probability of finding QDs with vanishing FSS is
general very small? (II) Why FSS and polarization angle differ dramatically
from QD to QD? and (III) Is there any direct connection between FSS,
polarization angle and morphology of QDs. These three questions are
essential for understanding optical properties of QDEs.

This Letter is devoted to address these three fundamental issues.  We propose an
effective model to describe the statistical features of FSS and polarization
angle and derive their corresponding distribution functions. We show that
their statistical properties can be fully characterized using at most three
independent measurable parameters. The effective model is then confirmed 
using atomistic pseudopotential calculations as well as experimental
measurements for several rather different types of QDEs. Potential
applications of the generic model is also discussed. These results yield a completely 
new physical picture for understanding optical properties of QDEs.

\textit{Theoretical Model.-} An analytical model which can capture the true symmetry
properties of QDs is required to better understand the statistical properties of QDEs.
To this end, the recently developed phenomenological model in Ref. [\onlinecite{Gong11}] is well fitted to this problem. From the
symmetry viewpoint, the Hamiltonian for a single QD can be written as $%
H=H_{2v}+V_{1}$, where $H_{2v}$ contains the kinetic energy and the
potential with the crystal $C_{2v}$ symmetry, and $V_{1}$ is the
perturbation potential with $C_{1}$ symmetry. We define two eigenvectors of $%
H_{2v}$ as $|3\rangle =|\Gamma _{2}+i\Gamma _{4}\rangle $ and $|4\rangle
=|\Gamma _{2}-i\Gamma _{4}\rangle $, which are exactly along either [110] or [1$%
\bar{1}$0] direction (ensured by the $C_{2v}$ symmetry) and also real
simultaneously (ensured by the time-reversal symmetry). An effective $%
2\times 2$ Hamiltonian can be constructed from these two bright states \cite%
{Gong11},
\begin{equation}
H=\bar{E}+\delta \sigma_{z}+\kappa \sigma_{x}  \label{Eq-H22},
\end{equation}%
where $\bar{E}+\delta =\langle 3|H|3\rangle $, $\bar{E}-\delta =\langle
4|H|4\rangle $, $\kappa =\langle 3|V_{1}|3\rangle $, and $\delta $, $\kappa $
$\in \mathbb{R}$. $\sigma_{x}$ and $\sigma_{z}$ are Pauli matrices. $\bar{E%
}$ defines the exciton energy, and $\Delta =2\sqrt{\kappa ^{2}+\delta ^{2}}$
defines the FSS. Physically, $\kappa $ describes the coupling between two
bright states, leading to the  deviation of the emission line from [110] and
[1$\bar{1}$0] directions. The wavefunction of the bright exciton can be written as $%
\psi =\cos (\theta )|3\rangle +\sin (\theta )|4\rangle $, where $\theta $ is the polarization angle with tan($%
\theta $) = $\kappa ^{-1}(\kappa \pm \sqrt{\delta ^{2}+\kappa ^{2}})$. Hence
$\delta =\pm \Delta \sin (2\theta )/2$, and $\kappa =\mp \Delta \cos
(2\theta )/2$.

The $V_{1}$ term can be uniquely determined with the following recipe. We
assume $V$ is the total potential of QDs (including all types of
interactions), and $G$ contains all irreducible representations of $C_{2v}$
point group. For any $g\in G$, $gH_{2v}g^{-1}=H_{2v}$, therefore $H_{2v}=T+(\sum_{g\in G}gV(\mathbf{r}%
)g^{-1})/|G|$, where $T$ is the kinetic energy and $|G|$ is the number of
symmetry operators. Then%
\begin{equation}
V_{1}(\mathbf{r})=V(\mathbf{r})-(\sum_{g\in G}gV(\mathbf{r})g^{-1})/|G|.
\label{Eq-V1}
\end{equation}%
Obviously, $\sum_{g}gV_{1}(\mathbf{r})g^{-1} \equiv 0$ for any $\mathbf{r}$.
The $C_{2v}$ symmetry ensures the direct connection between $\mathbf{r}$ and
$g\mathbf{r}g^{-1}$, whereas there are no correlations for other coordinate
pair ($\mathbf{r}$, $\mathbf{r}^{\prime })$ when $\mathbf{r}^{\prime }\neq g%
\mathbf{r}g^{-1}$. Thus the $V_{1}$ term, which depends essentially on the
local details of QDs, has the basic feature even in a single QD that the
potential should be spatially changed rapidly both in sign and magnitude.

The major difference between single QD and QDE is that in the latter case
the morphology variations of QDs make $H_{2v}$ and $V_{1}$ also varying. To
make the physical picture more transparent, we rewrite $H_{2v}=\bar{H}%
_{2v}+\delta H_{2v}$, where $\bar{H}_{2v}=\langle H_{2v}\rangle $, and $%
\delta H_{2v}$ defines the variations from QD to QD.  Now, $V_1$ can be treated
as an random potential from the viewpoint of QDEs due to its peculiar feature 
in Eq. \ref{Eq-V1}. The $\delta H_{2v}$, although
lacking similar feature, also exhibits some degrees of randomness. We
propose that the effective model in Eq. \ref{Eq-H22} for a single QD
is still valid to describe the optical properties of QDEs, with the
modification that $\delta $ and $\kappa$ be treated as independent random numbers
satisfying some particular distributions. Notice that both $%
\delta H_{2v}$ and $V_{1}$ contribute to the randomness of $\delta $, while
only $V_{1}$ contributes to $\kappa $. We expect $\langle \kappa \rangle =0$%
, but $\langle \delta \rangle \neq 0$ because QDs with $C_{2v}$ symmetry ($V_1 \equiv 0$) still have finite 
FSS. We set $\delta =\delta_{0}+\delta^{\prime }$, where $\delta _{0}=\langle \delta \rangle $. In the
following, it is essential to verify the randomness of $\delta $ and $\kappa
$ to validate our basic effective model. We will also show that $\delta _{0}$
characterizes the shape anisotropy effect of the QDEs.

\begin{figure}[tbp]
\centering
\includegraphics[width=3.0in]{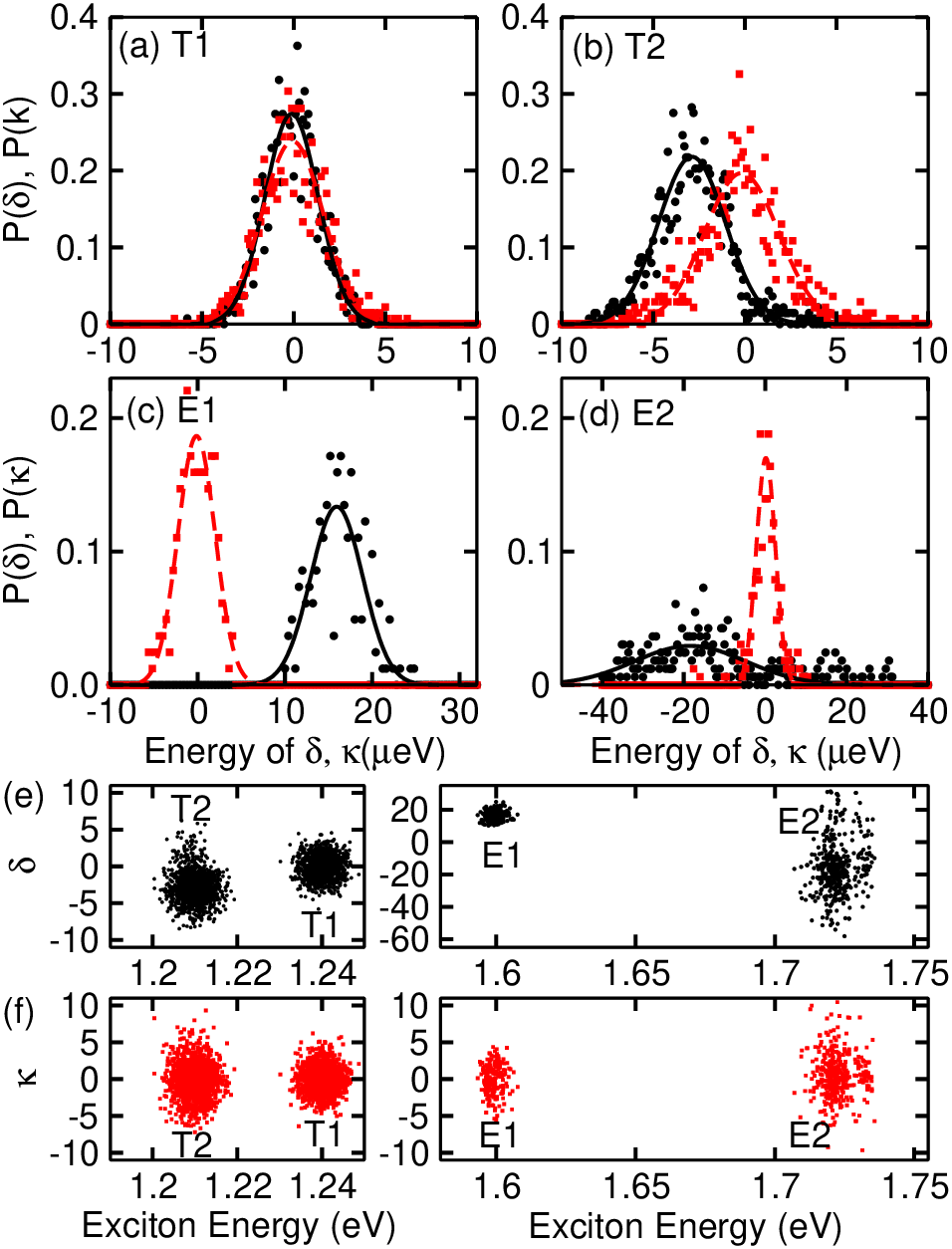}
\caption{(Color online). (a) - (d) distribution of $\protect\delta $ ($\bullet $) and $\protect\kappa
$ ($\blacksquare $) for different QDEs. The solid lines and dashed lines are the best fitting to Gaussian 
distributions, and the fitted parameters are tabled in Table \ref{table1}. The exciton energy dependence of $\protect%
\delta $ and $\protect\kappa $ are plotted in (e) and (f), respectively.}
\label{fig1}
\end{figure}

Intuitively, the distributions of FSS and polarization angle should depend
strongly on the morphology details of QDs, including size (base diameter, height),
shape, alloy profile, \textit{etc}. However, it is
impossible to quantitatively determine all these parameters in experiments\cite{NLiu,Denker,Walther}. 
We circumvent this difficult by defining several physical parameters
to fully characterize the statistical features of FSS and
polarization angle in QDEs. The distribution of any observable physical
quantity, say $f=f(\kappa ,\delta _{0},\delta ^{\prime })$, is defined as
\begin{equation}
P(z)=\int d\delta ^{\prime }d\kappa \delta (f-z)\mathcal{N}(\delta ^{\prime
},\sigma _{\delta})\mathcal{N}(\kappa ,\sigma _{\kappa }),
\label{eq-f}
\end{equation}%
where $\kappa $ and $\delta^{\prime }$ satisfy normal distributions $\mathcal{N}(x,\sigma )$,
with variations $\sigma_{\kappa} = \langle \kappa^2\rangle$ and $\sigma_{\delta} = \langle (\delta^{\prime})^2\rangle$, 
respectively. The exact distribution functions for FSS and polarization angle can be found in the supplementary material [\onlinecite{sup}]. 
The advantage of our strategy is that we are able to define and characterize the statistical features of FSS and 
polarization angle of the QDEs without knowing the morphology details of the samples. 

\begin{figure}[tbp]
\centering
\includegraphics[width=3.0in]{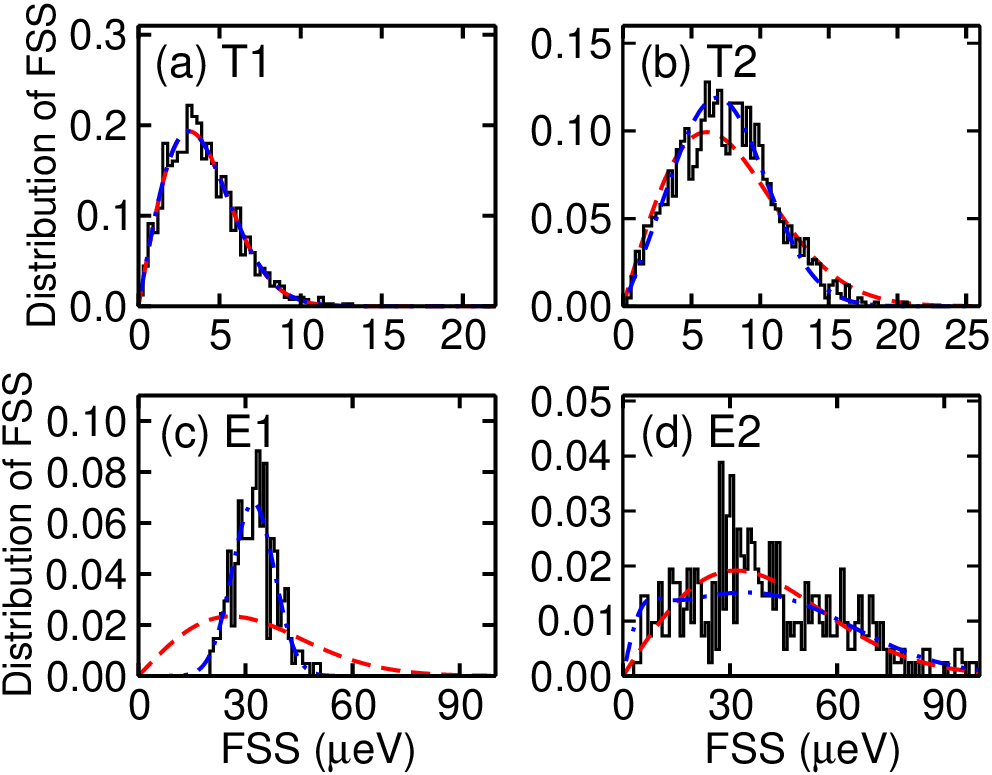}
\caption{(Color online). Distribution of FSS for different QDEs. In each panel, the folded
line represents the theoretical/experimental data, the dashed line is the best
fitting using Eq. \protect\ref{eq-goe}, and the dot-dashed line is from 
direct calculation using Eq. \protect\ref{eq-f} with the three parameters from Table \ref{table1}.}
\label{fig2}
\end{figure}

Here, we are particularly interested in the distribution of FSS with $\delta
_{0}=0$, which corresponds to the results of lens-shaped QDEs. The
distribution of FSS reads as
\begin{equation}
P(\Delta )={\frac{\Delta }{4\sigma _{\delta }\sigma _{\kappa }}}\exp
(-A_{+}\Delta ^{2})I_{0}(A_{-}\Delta ^{2}),  \label{eq-fssexact}
\end{equation}%
where $A_{\pm }=1/(16\sigma _{\kappa }^{2})\pm 1/(16\sigma_{\delta}^{2})$ and $I_{0}(x)$ is the modified Bessel function of the first kind.
We find that for a large class of QDEs the distribution of
FSS can be well approximated by the Wigner function \cite{RMatrix}
\begin{equation}
P(s)={\frac{\pi }{2}}s\exp (-{\frac{\pi }{4}}s^{2}),  \label{eq-goe}
\end{equation}%
with $s=\Delta /\langle \Delta \rangle $. Notice that Eq. \ref{eq-goe} is
the standard level spacing of the random matrix of Gaussian orthogonal
ensemble, in which all matrix elements are assumed to be
independently identically Gaussian distributed. Eq. \ref{eq-goe} exactly
describes the distribution of FSS only when $\delta _{0}=0$ and $\sigma
_{\kappa }=\sigma _{\delta}$. The above two equations clearly
demonstrate the strong repulsion between levels. More specifically, the
strong repulsion is related to the independence of off-diagonal and diagonal
elements, which leads to the appearance of the first $s$ term in \ref{eq-goe}. 
It is exactly this term that renders the
probability of finding QDs with vanishing FSS very small. It therefore
should be in stark contrast to the trivial random matrix without
off-diagonal elements where strong attractive between levels leads to level
spacings described by Poisson distribution \cite{RMatrix}.

The distribution of the polarization angle $\theta \in \lbrack -\pi /2,\pi
/2]$ at $\delta_0 = 0$ reads as \cite{ThetaNote}
\begin{equation}
P(\theta )={\frac{1}{\pi }}\cdot {\frac{1}{\eta \sin (2\theta )^{2}+\eta
^{-1}\cos (2\theta )^{2}}},  \label{eq-theta}
\end{equation}%
where $\eta =\sigma _{\delta }/\sigma _{\kappa }$. We see $P(\theta )=1/\pi $
when $\eta =1$. $P(\theta )$ reaches the maximum at $\theta =0,\pm \pi /2$
when $\eta >1$, or at $\theta =\pm \pi /4$ when $\eta <1$. The period of $%
P(\theta )$ is $\pi /2$. It is worth emphasizing that the fluctuation of $\theta$ 
is induced by the random coupling between the bright states from $V_1$, 
thus it is not an extrinsic effect\cite{Abb}. In the following, we will confirm the validity of the 
above analytical results using several rather different QDEs, some of which even have relative
large $\delta _{0}$. The validity of the above one-parameter equations will also be examined.

\begin{figure}[tbp]
\centering
\includegraphics[width=3.0in]{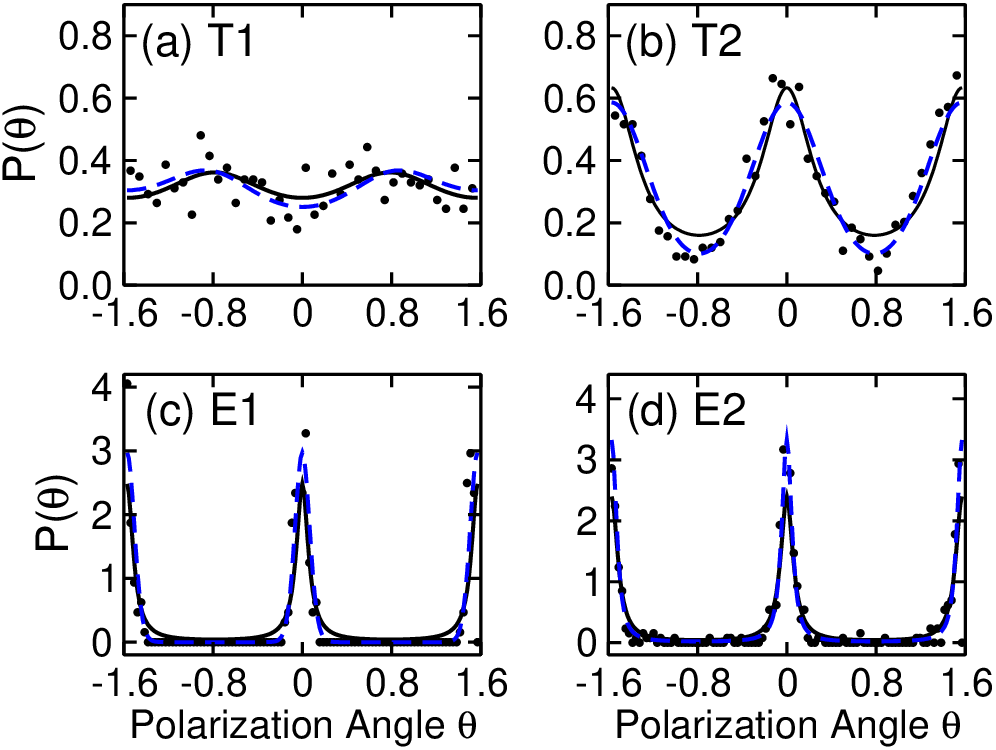}
\caption{(Color online). Distributions of polarization angles for different QDEs. In each
panel, the symbols represent the theoretical/experimental data, the solid line
is the best fitting using Eq. \protect\ref{eq-theta}, ad the dashed line
is from direct calculation using Eq. \protect\ref{eq-f} with the three parameters
from Table \ref{table1}.}
\label{fig3}
\end{figure}

In the limit of $\delta _{0}\gg \sigma _{\delta},\sigma _{\kappa }
$, the statistical features turn out to be very simple and can be understood
as follows. The polarization angle should be around either $\theta =0$ or $%
\theta =\pi /2$, thus $\eta \gg 1$. The FSS is most likely to be observed at
$\Delta \sim \delta _{0}$ in such types of QDEs, and the distribution of FSS
decays rapidly to zero when $|\Delta -\delta _{0}|\gg \sigma _{\delta}$. The distribution of FSS is more close to a Gaussian function
with width $\sqrt{\sigma _{\delta}^{2}+\sigma _{\kappa }^{2}}$.
This limit actually corresponds to QDEs with strong shape anisotropy.

\textit{Theoretical and experimental verifications.-} Above analytic results
are further confirmed using atomistic simulation and realistic experiments.
The simulation is based on the well-established atomistic pseudopotential method \cite{LWang1,LWang2,LWang3}. 
We consider three different In$_{x}$Ga$_{1-x}$As/GaAs QDEs: (T1) Lens QDs with fixed $%
x=0.6$, diameter $D=25$ nm, and height $h=3.5$ nm; (T2) Elongated QDs with $%
x=0.6$, diameter along [110] ([1$\bar{1}$0]) direction $D_{[110]}=26$ nm ($%
D_{[1\bar{1}0]}=24$ nm), $h=3.5$ nm; (T3) Lens QDs where $x$, $D$ and
$h$ are variables, with the mean values $\langle x\rangle =0.6$, $\langle
D\rangle =25$ nm, $\langle h\rangle =3.5$ nm, and variations up to 10\% of
the corresponding mean values. 
In the experiments we consider four different QDEs. The samples were grown by solid source
molecular beam epitaxy on GaAs (001) substrates. GaAs/AlGaAs QDs are obtained by infilling
self-assembled nanoholes fabricated in situ either by droplet etching\cite{Atkinson} or by selective 
AsBr$_3$ etching\cite{Rastelli}. The indium flushed self-assembled In(Ga)As/GaAs QDs are grown by 
Stranski-Krastanov  technique\cite{ZRW}. Excitons 
confined in the GaAs QD (QDE E1), In(Ga)As QD (QDE E4) and GaAs quantum well potential fluctuations 
(QDEs E2 and E3, respectively) are investigated. All the
microphotoluminescence spectroscopy measurements were performed at low temperature.
We study several rather different QDEs here to show the validity of the generic model. In the following, we choose
QDEs T1, T2, E1 and E2 to present our major findings, while all the
parameters for the QDEs, including the fitted results, are summarized in
Table \ref{table1} for comparison. More details can be found in the
supplementary material [\onlinecite{sup}].

\begin{table}[th]
\caption{Summarized parameters for different QDEs. The definition of QDs
can be found in the main text and supplementary material
\protect\cite{sup}. $N$ the sample volume of simulated or measured QDs with $%
E_{X}$ (eV) the mean exciton energy, and $\protect\sigma _{X}$ (meV) the
variation of the exciton energies. $\langle \protect\kappa \rangle $, $%
\protect\sigma _{\protect\kappa }$, $\protect\delta _{0}$, $\protect\sigma _{%
\protect\delta}$, $\langle \Delta \rangle $ (the mean value of
FSS) are all in unit of $\protect\mu $eV. $\protect\eta $ has been defined
in Eq. \protect\ref{eq-theta}. $\mathcal{G}^{(2)}=\langle (\protect\delta %
-\langle \protect\delta \rangle )(\protect\kappa -\langle \protect\kappa %
\rangle )\rangle /\protect\sigma_{\protect\delta}\protect\sigma %
_{\protect\kappa }$ measures the cross correlation between the two random
numbers. $\mathcal{P}$ (\%) defines the probability of finding QDs with FSS
smaller than the broadening of the emission line ($1$ $\protect\mu $eV).}%
\centering
\begin{tabular}{p{0.1\columnwidth}p{0.11\columnwidth}p{0.11\columnwidth}p{0.11\columnwidth}p{0.11\columnwidth}p{0.11\columnwidth}p{0.11\columnwidth}p{0.11\columnwidth}}
\hline\hline
QDs & T1 & T2 & T3 & E1 & E2 & E3 & E4 \\ \hline
$N$ & 1351 & 1381 & 7714 & 204 & 412 & 401  & 240 \\
$E_X$ & 1.240 & 1.210 & 1.186 & 1.598 & 1.722 & 1.768 & 1.386 \\
$\sigma_X$ & 2.9 & 3.0 & 28.3 & 2.5 & 3.6 & 5.1 & 4.4 \\
$\langle \kappa \rangle$ & -0.1 & -0.2 & -0.2 & -0.1 & 0.2 &  -0.3 & -0.4 \\
$\sigma_{\kappa}$ & 1.7 & 2.0 & 1.5 & 2.1 & 2.3 & 1.3  & 4.4 \\
$\delta_0 $ & -0.1 & -2.9 & 0.6 & 15.9 & -18.4 & -3.5 & -1.7 \\
$\sigma_{\delta}$ & 1.4 & 1.8 & 1.3 & 3.0 & 13.6  & 10.2 & 4.8 \\
$\langle \Delta \rangle$ & 4.0 & 7.7 & 3.8 & 32.6 & 39.7 & 15.4 & 11.7 \\
$\eta$ & 0.88 & 1.99 & 1.04 & 7.80 & 7.53  & 5.26 &1.0 \\
$\mathcal{G}^{(2)}$ & 0.002 & 0.003 & $\sim 0.0$ & -0.041 & -0.298 & 0.114  & -0.08 \\
$\mathcal{P}$ & 5.0 & 1.3 & 5.7 & $\sim 0.0$ & 0.2 & 0.1  & 0.5 \\
\hline\hline
\end{tabular}%
\label{table1}
\end{table}

For all QDEs we observe that the distributions of $\delta $ and $\kappa $
(see Fig. \ref{fig1}a -d) can be well fitted with Gaussian functions. We have
also confirmed that these random variables are independent of exciton energies
(see Fig. \ref{fig1}e, f). In QDE T1, we find $\delta_{0}\sim 0$ and $\sigma _{\delta}\simeq \sigma
_{\kappa }$, see Fig. \ref{fig1}a, which agree well with the experimental 
data in QDE E4; In elongated QDE T2 (Fig. \ref{fig1}b), the shape anisotropy leads to 
significant nonzero $\delta _{0}$. This observation qualitatively agrees with the results in 
experiments, see Fig. \ref{fig1}c-d, which have elongation either along [110] or [1$\bar{1}$0] 
direction. To verify the randomness of these two parameters, we calculate the
cross correlations between $\delta^{\prime}$ and $\kappa $ and find that
the correlation between them is indeed very small, see $\mathcal{G}^{(2)}$ in 
Table \ref{table1}. The experimental measured correlation is larger than our atomistic
simulation because much smaller sample volume is measured in experiments.
The fluctuations of $\delta $ and $\kappa $ both in simulations and
experiments are found to be of the order of several $\mu $eV; $\langle \kappa \rangle \sim 0$ 
is also consistent with expectation.

The connection between morphology and FSS can be established as follows. We
observe that the anisotropic effect arising from the lattice nonequivalence
of [110] and [1$\bar{1}$0] directions in zinc-blende crystals leads to a
fairly small $\delta _{0}$, whereas the anisotropic effect arising from
shape elongation generally leads to a significant $\delta _{0}$ (and hence
large FSS), as observed in QDE T2 and the experimental
samples. In experiments, the QDs exhibit apparent elongation either along
[110] or [1$\bar{1}$0] direction, and for the quantum well potential
fluctuation QDE, strong shape anisotropic effect is also expected\cite{Plumhof}. 
The basic conclusion that shape anisotropy has dominate contribution
to FSS is consistent with the recent reports by Plumhof\cite{Plumhof} and
Huo\cite{Huo}. However what we advance here is that the shape anisotropy can be
fully characterized by $\delta _{0}$. We note that the morphology of a
single QD is impossible to be precisely controlled in 
experiments. However, the statistical properties of QDEs maybe well controlled\cite{Malik}, 
therefore it provides an interesting arena in future to study the relationship between these
parameters and the growth environments, such as temperature, pressure, etc.

The distributions of FSS for QDEs are presented in Fig. \ref{fig2}. The
dashed lines are the best fitting using Eq. \ref{eq-goe}, while the
dot-dashed lines are calculated from Eq. \ref{eq-f} with the three
parameters obtained directly from fitting the distributions of $\delta $ and
$\kappa $ with Gaussian distributions (see Fig. \ref{fig1}). We have numerically confirmed that the distribution of FSS can be
well described using Eq. \ref{eq-goe} when $\delta _{0}<2\sigma _{\delta}$, and it is somewhat 
poorer for QDEs with large $\delta_{0}/\sigma _{\delta}$ (see QDE E1).The distributions of
polarization are presented in Fig. \ref{fig3} and generally much better agreement can
be obtained. The good agreement between analytical curves from Eq. \ref{eq-f}
and simulations/experiments clearly demonstrate the validity of our model.
Here we deliberately verify the validity of Eq. \ref{eq-goe} and \ref{eq-theta} 
to the condition $\delta _{0}\neq 0$ to provide important
reference for future researches in QDs and other nanostructures.

\textit{Discussions and concluding remarks.-} We now answer the three fundamental questions put forward in the 
introduction. Firstly, the probability of finding QDs with FSS smaller than 1 $\mu $eV is $\mathcal{P}%
=1/16\sigma _{\delta ^{\prime }}\sigma _{\kappa }$ when $\delta _{0}=0$. For
typical values of $\sigma_{\delta}$ and $\sigma _{\kappa }$, $%
\mathcal{P}\sim 1\%$. When $\delta_{0}\neq 0$, a prefactor $\exp
(-\delta _{0}^{2}/2\sigma_{\delta}^{2})$ arises,  %should be multiplied,
which further suppresses $\mathcal{P}$, see Table \ref{table1} (Q:I). Secondly, in a random potential $V_1$, it 
is possible to observe two QDs with the same exciton energies and the same FSSs, but fairly different polarization 
angles. No obvious correlation between FSS and polarization angle can be derived (Q:II). Finally, due to the randomness of $\delta $ and $\kappa $, the
morphology effect can be missed out in experiments of single QD\cite{Bester03, Bester05}. 
However, it can be recovered from experiments about QDEs. In particular, for QDE with a
large FSS ($\delta_0 \gg \sigma _{\delta}$, $\sigma _{\kappa }$),
the mean value of FSS itself can be used to characterize the shape anisotropy 
effect of QDE, in which condition the emissions should almost polarized along either [110] or [1$\bar{1}$0] direction (Q:III).

Several additional remarks are in order. Firstly, the basic idea can be
easily generalized to study the optical properties of high-symmetric QDEs.
Although the two bright states are degenerate for QDs with $C_{3v}$ or $%
D_{2d}$ symmetry\cite{Singh09}, the random term $V_{1}$ can still render the probability of
finding QDs with vanishing FSS small, as seen in recent experiments 
\cite{Mohan, Kuroda, Treu}. For high-symmetric QDEs, $\delta _{0}=0$, thus only
two independent parameters are required to fully characterize the
statistical properties of FSS and polarization angle. We estimate  $\sigma_{\kappa}$ and $\sigma _{\delta }$ 
$\sim $ 1 $\mu $eV using the results from Ref. [\onlinecite{Kuroda,Mohan}], which seems to a bit smaller
than that in $C_{2v}$ symmetric QDEs, see Table \ref{table1}. 
Secondly, the effective model is derived purely from symmetry argument, and is independent of the
morphology details of QDs, thus it is also applicable to study the optical
properties of other semiconductor nanostructures, e.g., quantum rod and 
colloid nanocrystals\cite{rod1, rod2, rod3}. As a generic feature, all the physical observations should 
exhibit some degree of random fluctuations\cite{Ming}. Thirdly, the leasing from QDE requires
the photons have good polarization property\cite{Saito, Jay}, and recently there are
indeed some remarkable progresses along this line \cite{Huffaker, HDjie, HLiu}. 
The statistical feature of polarization angle can find important
application in these fields. Finally, since the random potential $V_{1}$ is
impossible to be captured by any theoretical model, the theoretical modeling
can only be used to qualitatively, instead of quantitatively, interpret
the physical observations in experiments. To conclude, all these results and insights 
yield a completely new physical picture to understand the optical properties of QDEs.

\textit{Acknowledgement:} We thanks Armando Rastelli in JKU for valuable discussions, and also for his contribution 
in creating effective collaborations between the groups in USA, Germany and Austria. M. G. and C. Z. are supported by 
ARO (W911NF-12-1-0334, with part of the fund from DARPA-YFA), and NSF-PHY (1104546). J. L and A. Z are supported by 
the US Department of Energy, Office of Science, Basic Energy Sciences, Energy Frontier Research Centers, under Contract 
No. DE-AC36-08GO28308 to NREL. B. H., E. Z., R. T. and O. G. S. are partially supported by BMBF QuaHL-Rep (Contract No. 01BQ1032).


\begin{thebibliography}{99}
\bibitem{Akopian} N. Akopian, N. H. Lindner, E. Poem, Y. Berlatzky, J.
Avron, and D. Gershoni, B. D. Gerardot and P. M. Petroff, Phys. Rev. Lett.
\textbf{96}, 130501 (2006).

\bibitem{Stevenson} R. M. Stevenson, R. J. Young, P. Atkinson, K. Cooper, D.
A. Ritchie and A. J. Shields, Nature, \textbf{439}, 179 (2006).

\bibitem{Hofenbrak} R. Hafenbrak, S. M. Ulrich, P. Michler, L. Wang, A.
Rastelli and O. G. Schmidt, New J. Phys. \textbf{9}, 315 (2007).

\bibitem{Hudson} A. J. Hudson, R. M. Stevenson, A. J. Bennett, R. J. Young,
C. A. Nicoll, P. Atkinson, K. Cooper, D. A. Ritchie, and A. J. Shields,
Phys. Rev. Lett. \textbf{99}, 266802 (2007).

\bibitem{Young07} R. J Young, R. M. Stevenson, P. Atkinson, K. Cooper, D. A
Ritchie and A. J Shields, New J. Phys. \textbf{8}, 29 (2006).
%violation of Bell inequality

\bibitem{Young09} R. J. Young, R. M. Stevenson, A. J. Hudson, C. A. Nicoll,
D. A. Ritchie, and A. J. Shields, Phys. Rev. Lett. \textbf{102}, 030406
(2009).

\bibitem{Gong08} L. He, M. Gong, C.-F. Li, G.-C. Guo, and A. Zunger, Phys.
Rev. Lett. \textbf{101}, 157405 (2008).

\bibitem{WJP} J. Wang, M. Gong, G.-C. Guo, and L. He, Appl. Phys. Lett.
\textbf{101}, 063114 (2012).

\bibitem{Trotta} R. Trotta, E. Zallo, C. Ortix, P. Atkinson, J. D. Plumhof,
J. van den Brink, and O. G. Schmidt, Phys. Rev. Lett. \textbf{109}, 147401
(2012).

\bibitem{Bester03} G. Bester, S. Nair, and A. Zunger, Phys. Rev. B \textbf{67%
}, 161306 (2003).

\bibitem{Bester05} G. Bester, and A. Zunger, Phys. Rev. B \textbf{71},
045318 (2005).

\bibitem{Gong11} M. Gong, W. Zhang, G.-C. Guo, and L. He, Phys. Rev. Lett.
\textbf{106}, 227401 (2011).
% The following reference show how to tune the FSS using different methods

\bibitem{Kowalik} K. Kowalik, O. Krebs, A. Lemaitre, S. Laurent, P.
Senellart, P. Voisin, and J. Gaj, Appl. Phys. Lett. \textbf{86}, 041907
(2005).

\bibitem{Gerardot} B. D. Gerardot, S. Seidl, P. A. Dalgarno, R. J.
Warburton, D. Granados, J. M. Garcia, K. Kowalik, O. Krebs, K. Karrai, A.
Badolato, and P. M. Petroff, Appl. Phys. Lett. \textbf{90}, 041101 (2007).

\bibitem{Vogel} M. Vogel, S. Ulrich, R. Hafenbrak, P. Michler, L. Wang, A.
Rastelli, and O. Schmidt, Appl. Phys. Lett. \textbf{91}, 051904 (2007).

\bibitem{Marcet} S. Marcet, K. Ohtani, and H. Ohno, Appl. Phys. Lett.
\textbf{96}, 101117 (2010).

\bibitem{Luo12} J. W. Luo, R. Singh, A. Zunger, and G. Bester, Phys. Rev. B
\textbf{86}, 161302 (2012). %Zeeman field effect

\bibitem{Bennett} A. J. Bennett, M. A. Pooley, R. M. Stevenson, M. B. Ward,
R. B. Patel, A. B. de la Giroday, N. Skold, I. Farrer, C. A. Nicoll, D. A.
Ritchie, and A. J. Shields, Nature Phys. \textbf{6}, 947 (2010).

\bibitem{Stevenson06} R. M. Stevenson, R. J. Young, P. See, D. G. Gevaux, K.
Cooper, P. Atkinson, I. Farrer, D. A. Ritchie, and A. J. Shields, Phys. Rev.
B \textbf{73}, 033306 (2006).

\bibitem{Plumhof} J. D. Plumhof, V. Krapek, L. Wang, A. Schliwa, D. Bimberg,
A. Rastelli, and O. G. Schmid, Phys. Rev. B \textbf{81}, 121309(R) (2010).

\bibitem{Singh10} R. Singh, and G. Bester, Phys. Rev. Lett. \textbf{104}, 196803 (2010).

\bibitem{Seidl} S. Seidl, M. Kroner, A. Hagele, K. Karrai, R. J. Warburton,
A. Badolato, and P. M. Petroff, Appl. Phys. Lett. \textbf{88}, 203113 (2006).

\bibitem{Chris} Christopher E. Kuklewicz, Ralph N. E. Malein, Pierre M. Patroff, and Brian D. Gerardot, Nano Lett. {\bf 12}, 3761 (2012).

\bibitem{Luca} Luca Sapienza, Ralph N. E. Malein, Christopher E. Kuklewicz, Peter E. Kremer, Kartik Srinivasan, Andrew Griffiths, Edmund Clarke, Richard J. Warburton, Brian D. Gerardot, arXiv:1303.1122.

\bibitem{NLiu} N. Liu, J. Tersoff, O. Baklenov, A. L. Holmes, Jr., and C. K. Shih, Phys. Rev. Lett. \textbf{84}, 334 (2000).

\bibitem{Denker} U. Denker, M. Stoffel, and O. G. Schmidt, Phys. Rev. Lett. \textbf{90},  196102 (2003).

\bibitem{Walther} T. Walther, A. G. Cullis, D. J. Norris, and M. Hopkinson, Phys. Rev. Lett. \textbf{86}, 2381 (2001). 

\bibitem{sup} See supplemental material.

\bibitem{RMatrix} B. I. Shklovskii, B. Shapiro, B. R. Sears, P. Lambrianides, and H. B. Shore, Phys. Rev. B \textbf{47}, 11487 (1993).

\bibitem{Abb} M. Abbarchi, C. A. Mastrandrea, T. Kuroda, T. Mano, K. Sakoda, N. Koguchi, S. Sanguinetti, A. Vinattieri, and M. Gurioli, Phys. Rev. B \textbf{78}, 125321 (2008).

\bibitem{ThetaNote} Notice that we have taken the polarization angle of both 
bright states into account to symmetrize the distribution function, thus the distribution
of polarization angle is somewhat different from that reported by Plumhof\cite{Plumhof}. 

\bibitem{LWang1} L.-W. Wang, J. Kim, and A. Zunger, Phys. Rev. B \textbf{59}, 5678 (1999).

\bibitem{LWang2} L.-W. Wang and Alex Z., Phys. Rev. B \textbf{59}, 15806 (1999).

\bibitem{LWang3} A. J. Williamson, L.-W. Wang, and A. Zunger, Phys. Rev. B \textbf{62}, 12963 (2000).

\bibitem{Atkinson} P. Atkinson, E. Zallo, and O. G. Schmidt, J. Appl. Phys. {\bf 112}, 054303 (2012).

\bibitem{Rastelli} A. Rastelli, S. Stufler, A. Schliwa, R. Songmuang, C. Manzano, G. Costantini, K. Kern, 
A. Zrenner, D. Bimberg, and O. G. Schmidt, Phys. Rev. Lett. {\bf 92}, 166104 (2004).

\bibitem{ZRW} Z. R. Wasilewski,  S. Fafard, and J. P. McCaffrey, Journal of Crystal Growth. \textbf{201}, 1131 (1999).

\bibitem{Huo} H. Huo, A. Rastelli, and O. G. Schmidt, Appl. Phys. Lett. \textbf{102}, 152105 (2013).

\bibitem{Malik} Surama Malik, Christine Roberts, Ray Murray, and Malcolm Pate, Appl. Phys. Lett. \textbf{71}, 1987 (1997).

\bibitem{Singh09} R. Singh, and G. Bester, Phys. Rev. Lett. {\bf 103}, 063601 (2009).

\bibitem{Mohan} A. Mohan, M. Felici, P. Gallo, B. Dwir, A. Rudra, J. Faist,
and E. Kapon, Nature Photonics \textbf{4}, 302(2010).

\bibitem{Kuroda} T. Kuroda, T. Mano, N. Ha, H. Nakajima, H. Kumano, B. Urbaszek, M. Jo, M. Abbarachi, 
Y. Sakuma, K. Sakoda, I. Suemune, X. Marie, and T. Amand, arXiv:1302.6389.

\bibitem{Treu} J. Treu, C. Schneider,  A. Huggenberger, T. Braun, S. Reitzenstein, S. Hofling, and M. Kamp, Appl. Phys. Lett. \textbf{101}, 022102 (2012).

\bibitem{rod1} N. Le Thomas, E. Herz, O. Schöps, U. Woggon, and M. V. Artemyev, Phys. Rev. Lett. \textbf{94}, 016803 (2005). 

\bibitem{rod2} Qing Zhong Zhao, Peter A. Graf, Wesley B. Jones, Alberto Franceschetti, Jingbo Li, Lin-Wang, and Kwiseon Kim, Nano Lett. \textbf{7}, 3274 (2007).

\bibitem{rod3} S. P. Ahrenkiel, O. I. Micic, A. Miedaner, C. J. Curtis, J. M. Nedeljkovic, and A. J. Nozik, Nano Lett. \textbf{3}, 833 (2003).

\bibitem{Ming} Ming Gong, Junwei Luo {\it et al}, in prepare.

\bibitem{Saito}  Hideaki Saito, Kenichi Nishi, Shigeo Sugou, and Yoshimasa Sugimoto, Appl. Phys. Lett. \textbf{71}, 590 (1997).

\bibitem{Jay} P. Jayavel, H. Tanaka, T. Kita, O. Wada, H. Ebe, M. Sugawara, J. Tatebayashi, Y. Arakawa, Y. Nakata, and T. Akiyama, Appl. Phys. Lett. \textbf{84}, 1820 (2004).

\bibitem{Huffaker} D. L. Huffaker, G. Park, Z. Zou, O. B. Shchekin, and D. G. Deppe, Appl. Phys. Lett. \textbf{73}, 2564 (1998).

\bibitem{HDjie} H. S. Djie, B. S. Ooi, X.-M. Fang, Y. Wu, J. M. Fastenau, W. K. Liu, and M. Hopkinson, 
Optics Lett. \textbf{32}, 44 (2007).

\bibitem{HLiu} H. Y. Liu, S. L. Liew, T. Badcock, D. J. Mowbray, M. S. Skolnick, S. K. Ray, T. L. Choi, 
K. M. Groom, B. Stevens, F. Hasbullah, C. Y. Jin, M. Hopkinson, and R. A. Hogg, Appl. Phys. Lett. \textbf{89}, 073113 (2006).


\end{thebibliography}
\end{document}